\documentclass[sigconf]{acmart}

\usepackage{booktabs} 

\usepackage{subcaption}
\usepackage{enumitem}
\usepackage{url,bm}
\usepackage{latexsym}

\usepackage{algorithm}
\usepackage[noend]{algorithmic}
\usepackage{multirow}

\usepackage{etoolbox}
\newtoggle{conf} \toggletrue{conf}

\newcommand{\cut}[1]{}

\newcommand{\systemname}{{\em PriPeARL}}
\newcommand{\statType}{statType}
\newcommand{\entity}{e}
\newcommand{\secret}{s}
\newcommand{\dattr}{d_{attr}}
\newcommand{\dval}{d_{val}}
\newcommand{\atr}{T_a}
\newcommand{\laplacerandvar}{laplaceRandVar}
\newcommand{\timehier}{{\mathcal H}_{time}}
\newcommand{\enthierparam}{l}
\newcommand{\childrenentityset}{C_{\entity}}
\newcommand{\minthreshold}{\tau}
\newcommand{\trueCount}{trueCount}

\copyrightyear{2018} 
\acmYear{2018} 
\setcopyright{acmcopyright}
\acmConference[CIKM '18]{CIKM '18}{October 22--26, 2018}{Torino, Italy}
\acmPrice{15.00}
\acmDOI{10.1145/3269206.3272031}
\acmISBN{978-1-4503-6014-2/18/10}

\renewcommand\footnotetextcopyrightpermission[1]{}
\begin{document}
\title{PriPeARL: A Framework for Privacy-Preserving Analytics and Reporting at LinkedIn}

\author{Krishnaram Kenthapadi, \ Thanh T. L. Tran}
\orcid{1234-5678-9012}
\affiliation{  \institution{LinkedIn Corporation, USA}
}
\email{(kkenthapadi, tntran)@linkedin.com}

\begin{abstract}
Preserving privacy of users is a key requirement of web-scale analytics and reporting applications, and has witnessed a renewed focus in light of recent data breaches and new regulations such as GDPR. We focus on the problem of computing robust, reliable analytics in a privacy-preserving manner, while satisfying product requirements. We present \systemname, a framework for privacy-preserving analytics and reporting, inspired by differential privacy. We describe the overall design and architecture, and the key modeling components, focusing on the unique challenges associated with privacy, coverage, utility, and consistency. We perform an experimental study in the context of ads analytics and reporting at LinkedIn, thereby demonstrating the tradeoffs between privacy and utility needs, and the applicability of privacy-preserving mechanisms to real-world data. We also highlight the lessons learned from the production deployment of our system at LinkedIn.
\end{abstract}

\maketitle

\section{Introduction}\label{sec:intro}
\let\thefootnote\relax\footnote{~~~~ \\ $^*$ {\bf This paper has been accepted for publication in the 27th ACM International Conference on Information and Knowledge Management (CIKM 2018).} Both authors contributed equally to this work.}

Preserving privacy of users is a key requirement of web-scale data mining applications and systems such as web search, recommender systems, crowdsourced platforms, and analytics applications, and has witnessed a renewed focus in light of recent data breaches and new regulations such as GDPR~\cite{voigt2017eu}. As part of their products, online social media and web platforms typically provide different types of analytics and reporting to their users. 
For example, LinkedIn provides several analytics and reporting applications for its members as well as customers, such as ad analytics (key campaign performance metrics along different demographic dimensions), content analytics (aggregated demographics of members that viewed a content creator's article or post), and profile view statistics (statistics of who viewed a member's profile, aggregated along dimensions such as profession and industry). 
For such analytics applications, 
it is essential to preserve the privacy of members, since member actions could be considered as sensitive information. Specifically, we want to ensure that any one individual's action (e.g., click on an article or an ad) may not be inferred by observing the results of the analytics system. At the same time, we need to take into consideration various practical requirements for the associated product to be viable and usable.

In this paper, we investigate the problem of computing robust, reliable analytics in a privacy-preserving manner, while addressing product requirements such as coverage, utility, and consistency. We present \systemname, a framework for privacy-preserving analytics and reporting. We highlight the unique challenges associated with privacy, coverage, utility, and consistency while designing and implementing our system (\S\ref{sec:problem}), and describe the modeling components (\S\ref{sec:model}) and the system architecture (\S\ref{sec:arch}) to address these requirements. Our approach to preserving member privacy makes use of random noise addition inspired by differential privacy, wherein the underlying intuition is that the addition of a small amount of appropriate noise makes it harder for an attacker to reliably infer whether any specific member performed an action or not. 
Our system incorporates techniques such as deterministic pseudorandom noise generation to address certain limitations of standard differential privacy and performs post-processing to achieve data consistency.
We then empirically investigate the tradeoffs between privacy and utility needs using a web-scale dataset associated with LinkedIn Ad Analytics and Reporting platform (\S\ref{sec:exp}). We also highlight the lessons learned in practice from the production deployment of our system at LinkedIn (\S\ref{sec:lessons}). We finally discuss related work (\S\ref{sec:related}) as well as conclusion and future work (\S\ref{sec:conclusion}).

\section{Background and Problem Setting}\label{sec:problem}
We first provide a brief overview of analytics and reporting systems at LinkedIn, followed by a discussion of the key privacy and product requirements for such systems.

\subsection{Analytics and Reporting at LinkedIn}\label{sec:analyticsreporting}
Internet companies such as LinkedIn make use of a wide range of analytics and reporting systems as part of various product offerings. Examples include ad campaign analytics platform for advertisers, content analytics platform for content creators, and profile view analytics platform for members. The goal of these platforms is to present the performance in terms of member activity on the items (e.g., ads / articles and posts / member profile respectively), which can provide valuable insights for the platform consumers. For example, an advertiser could determine the effectiveness of an ad campaign across members from different professions, functions, companies, locations, and so on; a content creator could learn about the aggregated demographics of members that viewed her article or post; a member can find out professions, functions, companies, locations, etc. that correspond to the largest sources of her profile views. The platforms are made available typically as a web interface, displaying the relevant statistics (e.g., impressions, clicks, shares, conversions, and/or profile views, along with demographic breakdowns) over time, and sometimes also through corresponding APIs (e.g., ad analytics API). Figure~\ref{fig:adanalytics} shows a screenshot of LinkedIn's ad analytics and reporting platform (discussed in \S\ref{sec:expadanalytics}).

A key characteristic of these platforms is that they admit only a small number of predetermined query types as part of their user interface and associated APIs, unlike the standard statistical database setting that allows arbitrary aggregation queries to be posed. In particular, our analytics platforms allow querying for the number of member actions, for a specified time period, together with the top demographic breakdowns. We can abstractly represent the underlying database query form as follows.
\begin{itemize}
\item ``SELECT COUNT(*) FROM table(\statType, entity) WHERE timeStamp $\ge$ startTime AND timeStamp $\le$ endTime AND $\dattr = \dval$''
\end{itemize}
In the above query, table(\statType, entity) abstractly denotes a table in which each row corresponds to a member action (event) of statistics type, {\em \statType} for {\em entity} (e.g., {\em clicks} on a {\em given ad}), $\dattr$ the demographic attribute (e.g., title), and $\dval$ the desired value of the demographic attribute (e.g., ``Senior Director''). In practice, these events could be preprocessed and stored in a partially aggregated form so that each row in the table corresponds to the the number of actions (events) for a ({\em \statType}, {\em entity}, $\dattr$, $\dval$, the most granular time range) combination, and the query computes the sum of the number of member actions satisfying conditions on the desired time range and the demographic attribute-value pair.

\subsection{Privacy Requirements}\label{sec:privacyrequirements}
We next discuss the requirement for preserving the privacy of LinkedIn members. Our goal is to ensure that an attacker cannot infer whether a member performed an action (e.g., click on an article or an ad) by observing the results shown by the analytics and reporting system, possibly over time. We assume that the attacker may have knowledge of attributes associated with the target member (e.g., obtained from this member's LinkedIn profile) as well as knowledge of all other members that performed similar action (e.g., by creating fake accounts that the attacker has then control over).

At first, it may seem that the above assumptions are strong, and the aggregate analytics may not reveal information about any member's action. However, we motivate the need for such privacy requirements by illustrating potential attacks in the context of ad analytics. Consider a campaign targeted to ``Senior directors in US, who studied at Cornell.'' As such a campaign is likely to match several thousands of members, it will satisfy any minimum targeting threshold and hence will be deemed valid. However, this criterion may match exactly one member within a given company (whose identity can be determined from the member's LinkedIn profile or by performing search for these criteria), and hence the company-level demographic breakdowns of ad clicks could reveal whether this member clicked on the ad or not. A common approach to reducing the risk of such attacks is to use a (deterministic) minimum threshold prior to showing the statistics. However, given any fixed minimum threshold $k$, the attacker can create $k-1$ or more fake accounts that match the same criteria as the target member, and have these accounts click on the ad so that the attacker can precisely determine whether the member clicked on the ad from the company-level ad click count. A larger fixed threshold would increase the effort involved in this attack, but does not prevent the attack itself. 

Similarly, we would like to provide incremental privacy protection, that is, protect against attacks based on incremental observations over time. We give an example to demonstrate how, by observing the reported ad analytics over time, a malicious advertiser may be able to infer the identity of a member that clicked on the ad. Consider an ad campaign targeted to ``all professionals in US with skills, `leadership' and `management' and at least 15 years of experience.'' Suppose that this ad receives a large number of clicks from leadership professionals across companies initially, and afterwards, on a subsequent day, receives just one click causing the ad click breakdowns for `title = CEO' and `company = LinkedIn' to be incremented by one each. By comparing these reported counts on adjacent days, the advertiser can then conclude that LinkedIn's CEO clicked on the ad.

The above attacks motivate the need for applying rigorous techniques to preserve member privacy in analytics applications, and thereby not reveal exact aggregate counts. However, we may still desire utility and \cut{different types of }data consistency, which we discuss next.

\subsection{Key Product Desiderata}\label{sec:requirements}

\subsubsection{Coverage and Utility}
It is desirable for the aggregate statistics to be available and reasonably accurate for as many action types, entities, demographic attribute/value combinations, and time ranges as possible for the analytics and reporting applications to be viable and useful.

\subsubsection{Data Consistency}

We next discuss the desirable properties for an analytics platform with respect to different aspects of data consistency for the end user, especially since the platform may not be able to display true counts due to privacy requirements. We note that some of these properties may not be applicable in certain application settings, and further, we may choose to either partially or fully sacrifice certain consistency properties either to achieve better privacy and/or utility. We discuss such design choices in \S\ref{sec:model}, \S\ref{sec:exp}, and \S\ref{sec:lessons}.

{\em Consistency for repeated queries (C1)}: The reported answer should not change when the same query is repeated (assuming that the true answer has not changed). For example, the reported number of clicks on a given article on a fixed day in the past should remain the same when queried subsequently. We treat this property as an essential one.

{\em Consistency over time (C2)}: The combined action counts should not decrease over time. For example, the reported total number of clicks on an article by members satisfying a given predicate at time $t_1$ should be at most that at time $t_2$ if $t_1 < t_2$.

{\em Consistency between total and breakdowns (C3)}: The reported total action counts should not be less than the sum of the reported breakdown counts. For example, the displayed total number of clicks on an article cannot be less than the sum of clicks attributed to members from different companies. 
We do not require an equality check since our applications typically report only the top few largest breakdown counts as these provide the most valuable insights about the members engaging with the product.

{\em Consistency across entity hierarchy (C4)}: When there is a hierarchy associated with the entities, the total action counts for a parent entity should be equal to the sum of the action counts over the children entities. For example, different ads could be part of the same campaign, different campaigns could be part of a campaign group, and several campaign groups could be part of an advertiser's account.

{\em Consistency across action hierarchy (C5)}: When there is a hierarchy associated with actions such that a parent action is a prerequisite for a child action (e.g., an article would need to be impressed (shown) to the member, before getting clicked), the count for the parent action should not be less than the count for the child action (e.g., the number of impressions cannot be less than the number of clicks).

{\em Consistency for top $k$ queries (C6)}: The top $k$ results reported for different choices of $k$ should be consistent with each other. For example, the top 10 titles and the top 5 titles respectively of members that clicked on an article should agree on the first 5 results.

\subsection{Problem Statement}\label{sec:problemstmt}
Our problem can thus be stated as follows: {\em How do we compute robust, reliable analytics in a privacy-preserving manner, while addressing the product desiderata such as coverage, utility, and consistency? How do we design the analytics computation system to meet the\cut{ immediate and future} needs of LinkedIn products?} We address these questions in \S\ref{sec:model} and \S\ref{sec:arch} respectively.
 \section{Privacy Model and Algorithms}\label{sec:model}
We present our model and detailed algorithm for achieving privacy protection in an analytics and reporting setting.
Our approach modifies the reported aggregate counts using a random noise addition mechanism, inspired by differential privacy~\cite{DKM+06,DMNS06}.
Differential privacy is a formal guarantee for preserving the privacy of any individual when releasing aggregate statistical information about a set of people. In a nutshell, the differential privacy definition requires that the probability distribution of the released results be nearly the same irrespective of whether an individual's data is included as part of the dataset. As a result, upon seeing a published statistic, an attacker would gain very little additional knowledge about any specific individual.

\begin{definition} \cite{DMNS06} A randomized mapping ${\mathcal K}$ satisfies $\epsilon$-differential privacy if for all pairs of datasets ($D, D'$) differing in at most one row, and all $S \subseteq Range({\mathcal K})$,
\[ Pr[{\mathcal K}(D) \in S] \le e^{\epsilon} \cdot Pr[{\mathcal K}(D') \in S], \]
where the probability is over the coin flips of ${\mathcal K}$.
\end{definition}

Formally, this guarantee is achieved by adding appropriate noise (e.g., from Laplace distribution) to the true answer of a statistical query function (e.g., the number of members that clicked on an article, or the histogram of titles of members that clicked on an article), and releasing the noisy answer. The magnitude of the noise to be added depends on the $L_1$ sensitivity of the query (namely, the upper bound on the extent to which the query output can change, e.g., when a member is added to or removed from the dataset), and the desired level of privacy guarantee ($\epsilon$).

\begin{definition} \cite{DMNS06} The $L_1$ sensitivity of a query function, $f: {\mathcal D} \rightarrow {\mathbf R}^d$ is defined as $\Delta(f) = \max_{D, D'} || f(D) - f(D')||_1$ for all pairs of datasets ($D, D'$) differing in at most one row.
\end{definition}

\begin{theorem} \cite{DMNS06} Given a query function $f: {\mathcal D} \rightarrow {\mathbf R}^d$, the randomized mechanism ${\mathcal K}$ that adds noise drawn independently from the Laplace distribution with parameter $\frac{\Delta(f)}{\epsilon}$ to each of the $d$ dimensions of $f(D)$ satisfies $\epsilon$-differential privacy.
\end{theorem}

For our application setting, we adopt {\em event-level} differential privacy~\cite{DNPR10}, in which the privacy goal is to hide the presence or absence of a single event, that is, any one action of any member. 
Under this notion, the sensitivity for the query shown in \S\ref{sec:analyticsreporting} equals 1.

We next describe our approach for adding appropriate random noise to demographic level analytics, and for performing post-processing to achieve different levels of consistency. We first present an algorithm for generating pseudorandom rounded noise from Laplace distribution for a given query (Algorithm~\ref{alg:prln}), followed by an algorithm for computing noisy count for certain canonical queries (Algorithm~\ref{alg:cnc}), and finally the main algorithm for privacy-preserving analytics computation (Algorithm~\ref{alg:nc}), which builds on the first two algorithms.

\subsection{Pseudorandom Laplace Noise Generation}\label{sec:prln}
A key limitation with the standard differential privacy approach is that the random noise can be removed, by issuing the same query many times, and computing the average of the answers. Due to this reason and also for ensuring consistency of the answer when the same query is repeated (e.g., the advertiser returning to check the analytics dashboard with the same filtering criteria), we chose to use a deterministic, pseudorandom noise generation algorithm. The idea is that the noise value chosen for a query is fixed to that query, or the same noise is assigned when the same query is repeated.

Given the statistical query, the desired privacy parameter, and the fixed secret, we generate a (fixed) pseudorandom rounded noise from the appropriate Laplace distribution using Algorithm~\ref{alg:prln}. First, the secret and the query parameters are given as input to the deterministic function, {\em GeneratePseudorandFrac}, which returns a pseudorandom fraction between 0 and 1. Treating this obtained fraction as sampled from the uniform distribution on $(0,1)$, we apply the inverse cumulative distribution function (cdf) for the appropriate Laplace distribution to get the pseudorandom noise. Finally, we round the noise to the nearest integer since it is desirable for the reported noisy counts to be integers.

The function, {\em GeneratePseudorandFrac}, can be implemented in several ways. One approach would be to concatenate the query parameters and the fixed secret, then apply a cryptographically secure hash function (e.g., SHA-256), and use the hash value as the seed to a pseudorandom number generator that gives a pseudorandom fraction uniformly distributed in between 0 and 1. To protect against length extension attack and potential collisions, it may be desirable to avoid keyed hashing and instead use a cryptographically secure and unbiased hash function such as HMAC with SHA-256 (HMAC-SHA256)~\cite{bellare1996keying}. This factor needs to be weighed against the computational efficiency requirements, which could favor simpler implementations such as applying a more efficient hash function and scaling the hash value to $(0,1)$ range, treating the hash value to be a uniformly distributed hexadecimal string in its target range. Note that the fixed secret is used so that an attacker armed with the knowledge of the algorithm underlying {\em GeneratePseudorandFrac} as well as the query parameters would not be able to compute the noise value.

\begin{algorithm}
\caption{Pseudorandom Laplace Noise Generation Algorithm, {\em GeneratePseudorandomLaplaceNoise}}\label{alg:prln}
\begin{algorithmic}[1]
  \STATE \textbf{Inputs}: 
  Fixed secret, $\secret$;
  Statistics type, $\statType$;
  Entity Id, $\entity$;
  Demographic (attribute, value) pair, $(\dattr, \dval)$;
  Atomic time range, $\atr$;
  Privacy parameter, $\epsilon$.
  \STATE \textbf{Output}: Corresponding pseudorandom Laplace noise value.
  \STATE $p := GeneratePseudorandFrac(\secret, \statType, \entity, \dattr, \dval, \atr)$
  \STATE $\laplacerandvar := \frac{-1}{\epsilon} \textrm{sgn}(p - 0.5) \ln(1 - 2 |p - 0.5|)$
  \RETURN round($\laplacerandvar$)
\end{algorithmic}
\end{algorithm}

\subsection{Canonical Noisy Count Computation}\label{sec:cnc}
We compute the noisy count for a canonical statistical query using Algorithm~\ref{alg:cnc}. A canonical query takes the form shown in \S\ref{sec:analyticsreporting}, wherein the startTime and the endTime are constrained to map to an atomic time range (discussed in \S\ref{sec:nc}). The algorithm computes the true answer for the query by probing the underlying database, and then adds a fixed, rounded, pseudorandom Laplace noise by invoking the function, {\em GeneratePseudorandFrac} (\S\ref{sec:prln}). In case the noisy answer is negative, a count of zero is reported instead to ensure consistency over time. In this manner, we ensure that the combined action counts do not decrease over time since a query over a longer time range could then be broken into canonical queries, whose results could be summed up.

\begin{algorithm}
\caption{Canonical Noisy Count Computation Algorithm, {\em ComputeCanonicalNoisyCount}}\label{alg:cnc}
\begin{algorithmic}[1]
  \STATE \textbf{Inputs}: 
  Fixed secret, $\secret$;
  Statistics type, $\statType$;
  Entity Id, $\entity$;
  Demographic (attribute, value) pair, $(\dattr, \dval)$;
  Atomic time range, $\atr$;
  Privacy parameter, $\epsilon$.
  \STATE \textbf{Output}: Noisy demographic level statistics value for the canonical query.
  \STATE Compute true count, $\trueCount$ of $\statType$ for entity $\entity$ over the atomic time range $\atr$ for the demographic (attribute, value) pair, $(\dattr, \dval)$.
  \STATE $noise := GeneratePseudorandomLaplaceNoise(\secret, \statType, \entity,$ $(\dattr, \dval), \atr, \epsilon)$
  \RETURN $\max(\trueCount + noise, 0)$
\end{algorithmic}
\end{algorithm}

\subsection{Privacy-preserving Analytics Computation}\label{sec:nc}
We next present our main algorithm for computing privacy-preserving analytics in Algorithm~\ref{alg:nc}. This algorithm computes the answer to an arbitrary statistical query of the form shown in \S\ref{sec:analyticsreporting}, achieving a balance between privacy, consistency, and utility needs.

We first discuss the special handling for the condition stated in line 3 of Algorithm~\ref{alg:nc}. If the entity is at a broader level and consists of very few children entities (e.g., an ad campaign id that corresponds to just 1 or 2 ads), we would like to provide consistency across entity hierarchy (C4). The underlying rationale is that discrepancy between the reported statistics for parent and children entities would cause a poor experience in extreme cases of the parent containing just one child or very few children. For instance, an advertiser who creates a campaign with just one ad may even perceive such discrepancy as a bug. However, this difference becomes less pertinent as the number of children increases. Hence, our algorithm recursively sums up the noisy counts for the children entities when the number of children is at most the entity hierarchy consistency parameter, $\enthierparam$ (given as input). As $\enthierparam$ increases, we satisfy C4 to a greater extent, although at the cost of reduced utility (noise with larger variance is added) and increased latency (due to a fan-out of recursive calls, possibly across multiple levels in the entity hierarchy).

If the condition discussed above is not satisfied, the algorithm first partitions the input time range into a minimal set, $\Delta$ of {\em atomic time ranges}, obtains the noisy counts for each time range in $\Delta$ using Algorithm~\ref{alg:cnc}, and computes their sum (lines 6-7). Given a fixed hierarchy of time ranges, we define an {\em atomic time range} to be a range that exactly maps to some level in the hierarchy. For example, let the hierarchy be specified as 3-hour epochs beginning at 12am, 3am, 6am, $\ldots$ $\leftarrow$ day  
$\leftarrow$ month $\leftarrow$ quarter $\leftarrow$ year. 
Then, using M/D HH:MM notation and assuming the same year, (1/1 15:00, 1/1 18:00), (1/1 00:00, 1/2 00:00), and (1/1 00:00, 4/1 00:00) are examples of valid atomic time ranges, while (1/1 15:00, 1/1 21:00), (1/1 00:00, 1/3 00:00),  and (3/31 21:00, 8/2 03:00) are not. 
The range, (3/31 21:00, 8/2 03:00) can be minimally partitioned into the following: 
(3/31 21:00, 4/1 00:00) [3-hour epoch], 
(4/1 00:00, 7/1 00:00) [quarter],
(7/1 00:00, 8/1 00:00) [month],
(8/1 00:00, 8/2 00:00) [day], and
(8/2 00:00, 8/2 03:00) [3-hour epoch].

\begin{algorithm}
\caption{Privacy-preserving Analytics Computation Algorithm, {\em ComputeNoisyCount}}\label{alg:nc}
\begin{algorithmic}[1]
  \STATE \textbf{Inputs}: 
  Fixed secret, $\secret$;
  Statistics type, $\statType$;
  Entity Id, $\entity$;
  Demographic (attribute, value) pair, $(\dattr, \dval)$;
  Time range, $T$;
  Privacy parameter, $\epsilon$;
  Minimum threshold, $\minthreshold$;
  Entity hierarchy consistency parameter, $\enthierparam$;
  Hierarchy of time ranges, $\timehier$.
  \STATE \textbf{Output}: Noisy demographic level statistics value.
  \IF {entity $\entity$ is at a broader level, and corresponds to a set $\childrenentityset$ of at most $\enthierparam$ children entities}
  \RETURN $\sum_{\textrm{entity, } f \in \childrenentityset} ComputeNoisyCount(\secret, \statType, f,$ $(\dattr, \dval), T, \epsilon, \minthreshold, \enthierparam, \timehier)$
  \ELSE
    \STATE Partition $T$ into a minimal set $\Delta$ of atomic time ranges based on the time range hierarchy $\timehier$.
    \STATE $noisyCount := \sum_{\atr \in \Delta} ComputeCanonicalNoisyCount(\secret,$ $\statType,$ $\entity, (\dattr, \dval), \atr, \epsilon)$
    \IF {$noisyCount < \minthreshold$}
      \STATE $noisyCount := 0$
    \ENDIF
    \RETURN $noisyCount$
  \ENDIF
\end{algorithmic}
\end{algorithm}

The above partition-based approach is chosen for the following reasons. First, it provides better utility for longer time range queries (by adding noise with smaller variance) while partially sacrificing consistency over time (C2), as discussed more in \S\ref{sec:lessons}. Second, it helps with privacy guarantees by bounding the number of time range queries involving a given member action event. For a given demographic attribute/value predicate and entity, each event would only be part of at most as many queries as levels in the time range hierarchy. In this context, the partitioning of the time range into a minimal set of atomic time ranges can be thought of as analogous to the binary counting mechanism proposed in~\cite{CSS11}. Finally, our partitioning approach protects against the time range split-based averaging privacy attack in which an attacker could average out the noise for a given time range query by splitting the time range into two halves at different intermediate points, obtaining the sum of the noisy counts for each such pair of queries, and then averaging these sums to reliably estimate the true count for the original time range query (this is similar to the split averaging attack in~\cite{francis2017diffix}).

Before returning the result, we apply a minimum threshold based post-processing step: if the noisy count is below the given minimum threshold parameter, $\minthreshold$, we report a count of zero instead (lines 8-9). As discussed in \S\ref{sec:lessons}, we suppress small counts as these may not have sufficient validity. We can achieve a balance between privacy, validity, and product coverage needs by varying the privacy parameter, $\epsilon$ and the threshold parameter, $\minthreshold$: If stronger privacy is desired, we could choose a smaller $\epsilon$, and thereby add noise with a larger variance, due to which small counts may have lesser validity. We could then opt to suppress them using a larger $\minthreshold$, thereby reducing product coverage (fewer demographic breakdowns).

We can derive the effective privacy guarantee assuming that noisy answers to every possible canonical query is published. Algorithm~\ref{alg:nc} can then be viewed as post-processing this published dataset, which does not incur further differential privacy loss. Denote the number of demographic attributes of interest by $n_{attr}$, the number of levels in the time range hierarchy by $n_{time}$, and the number of levels in the entity hierarchy by $n_{ent}$. Each member action (e.g., a member clicking on an ad) contributes to at most $n_{attr} \cdot n_{time} \cdot n_{ent}$ canonical queries, and hence the overall event-level differential privacy loss can be shown to be bounded by $n_{attr} \cdot n_{time} \cdot n_{ent} \cdot \epsilon$ using composition theorem~\cite{dwork2014algorithmic}. 
Note that such a theoretical guarantee is under worst case assumptions, and may not be meaningful
in practice (e.g., as large as 36 for six demographic attributes ($n_{attr} = 6$), three time range levels of 3-hour epoch $\leftarrow$ day $\leftarrow$ month ($n_{time} = 3$), and two entity levels (e.g., campaign and account, $n_{ent} = 2$) with $\epsilon = 1$). In addition, this guarantee is for {\em event-level}, and not for {\em user-level} differential privacy.

\subsection{Discussion of Consistency Checks\cut{ and Other Practical Considerations}}
Depending on the specific analytics or reporting application, we may apply post-processing steps to ensure consistency, and other algorithmic modifications.

{\em Consistency for repeated queries (C1)}: is always ensured due to the application of pseudorandom noise.

{\em Consistency over time (C2)}: The thresholding step in Algorithm~\ref{alg:cnc} is performed towards achieving C2, although C2 may be violated when the time range partitioning proceeds to a broader atomic time range in the hierarchy (e.g., the addition of the last day of March could introduce a quarter in the place of months/days, causing potential violation of C2). 

{\em Consistency between total and breakdowns (C3)}: In many practical analytics settings, there would be a long tail of breakdowns not captured by the top few values that are displayed in the application. Consequently, the sum of the top few breakdown values is typically considerably less than the total count, which is likely to be preserved even with noise addition. Further, in certain cases, a member could be associated with multiple values for the same attribute (e.g., a member with the job title, `Founder and CEO' could belong to both `Founder' and `CXO' seniority levels), and hence C3 may not hold even for true counts.

{\em Consistency across entity hierarchy (C4)}: As discussed in \S\ref{sec:nc}, C4 is partially satisfied.

{\em Consistency across action hierarchy (C5)}: We chose to not focus on C5 for two reasons: (1) As the count for a parent action is typically at least an order of magnitude times the count for a child action (e.g., far more impressions than clicks), noise addition is less likely to have an effect on C5. (2) Due to possible time delays between parent and child actions (e.g., a conversion could happen a few days after an ad click), it is possible for C5 to not hold for a given reporting period.

{\em Consistency for top $k$ queries (C6)}: If we compute the entire histogram of counts for a given attribute, C6 would be satisfied since we would be using the same ranked list of all values based on the noisy counts. However, when the number of possible values is large (e.g., there are hundreds of thousands of companies and over 25K titles), computing the entire noisy histogram could be computationally expensive. One potential heuristic is to obtain the top $k_{max}$ values using true counts (thereby deviating from differential privacy guarantees) for a sufficiently large $k_{max}$, and then reorder after adding noise, thereby satisfying C6 for all $k \le k_{max}$.

 \section{Privacy-preserving Analytics System Design and Architecture}\label{sec:arch}
\begin{figure}[t]
\centering
\includegraphics[width=0.5\textwidth]{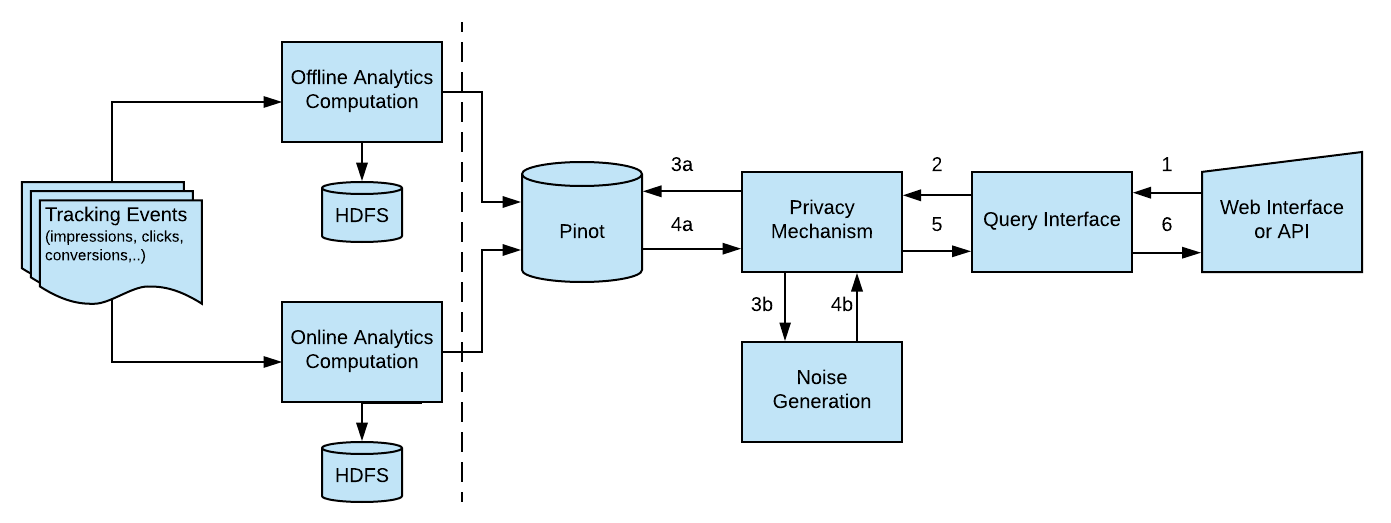}
\caption{Privacy-preserving Analytics System Architecture}
\label{fig:privacyarch}
\end{figure}

We describe the overall design and architecture of the privacy-preserving analytics computation system deployed as part of LinkedIn products. Our system consists of an online component that provides an interactive interface (or API) for presenting different types of analytics about member actions, and an offline/nearline component for generating the necessary tables containing the granular member actions (events). Figure~\ref{fig:privacyarch} presents the key components of our system.

{\em Analytics Computation Workflows}: 
The tables needed for computing different analytics about member actions are stored as part of LinkedIn's Pinot system. Pinot is a realtime distributed OLAP datastore, designed to deliver realtime analytics with low latency in a scalable manner~\cite{LinkedInPinot}. Two workflows are set up to process raw tracking data (in the form of various Kafka~\cite{kreps2011kafka} events generated from the member facing application, such as impressions as well as actions including clicks and conversions) and compute partially aggregated analytics, which are combined and ingested into the Pinot system. The first workflow is run daily in an offline fashion to compute the analytics for the previous day and persist to the datastore. The second workflow is run every few hours in an `online' fashion to compute intra-day incremental changes. The data pushed to Pinot contains several key fields including the granular time range, entity (and its hierarchy, if applicable), demographic attribute and value, and impression counts and different types of action counts. This data is fine-grained, and is aggregated corresponding to the analytics request arising from the web interface or the API.

{\em Online System for Presenting Analytics}:
Our online system uses a service oriented architecture for retrieving and presenting privacy-preserving analytics about member actions corresponding to the request from the user facing product (web interface) or API. First, this request is issued to the `Query Interface' component (step 1), which processes and transforms the request into underlying database queries, which are then passed to the `Privacy Mechanism' component (step 2). This component obtains the true answers to these queries by probing the Pinot system (steps 3a \& 4a), generates the appropriate noise (steps 3b \& 4b), performs post-processing consistency checks following Algorithm~\ref{alg:nc}, and returns the set of noisy counts (step 5), which are presented to the calling service (step 6).
 \section{Empirical Evaluation}\label{sec:exp}
We next present an empirical evaluation of our system, \systemname, for computing privacy-preserving analytics for the LinkedIn Ad Analytics \& Reporting platform.

\subsection{LinkedIn Ad Targeting \& Analytics Overview}\label{sec:expadanalytics}

We first give an overview of ad targeting and analytics at LinkedIn, which sets the context for our experiments.

\begin{figure}[t]
\centering
\includegraphics[width=0.5\textwidth]{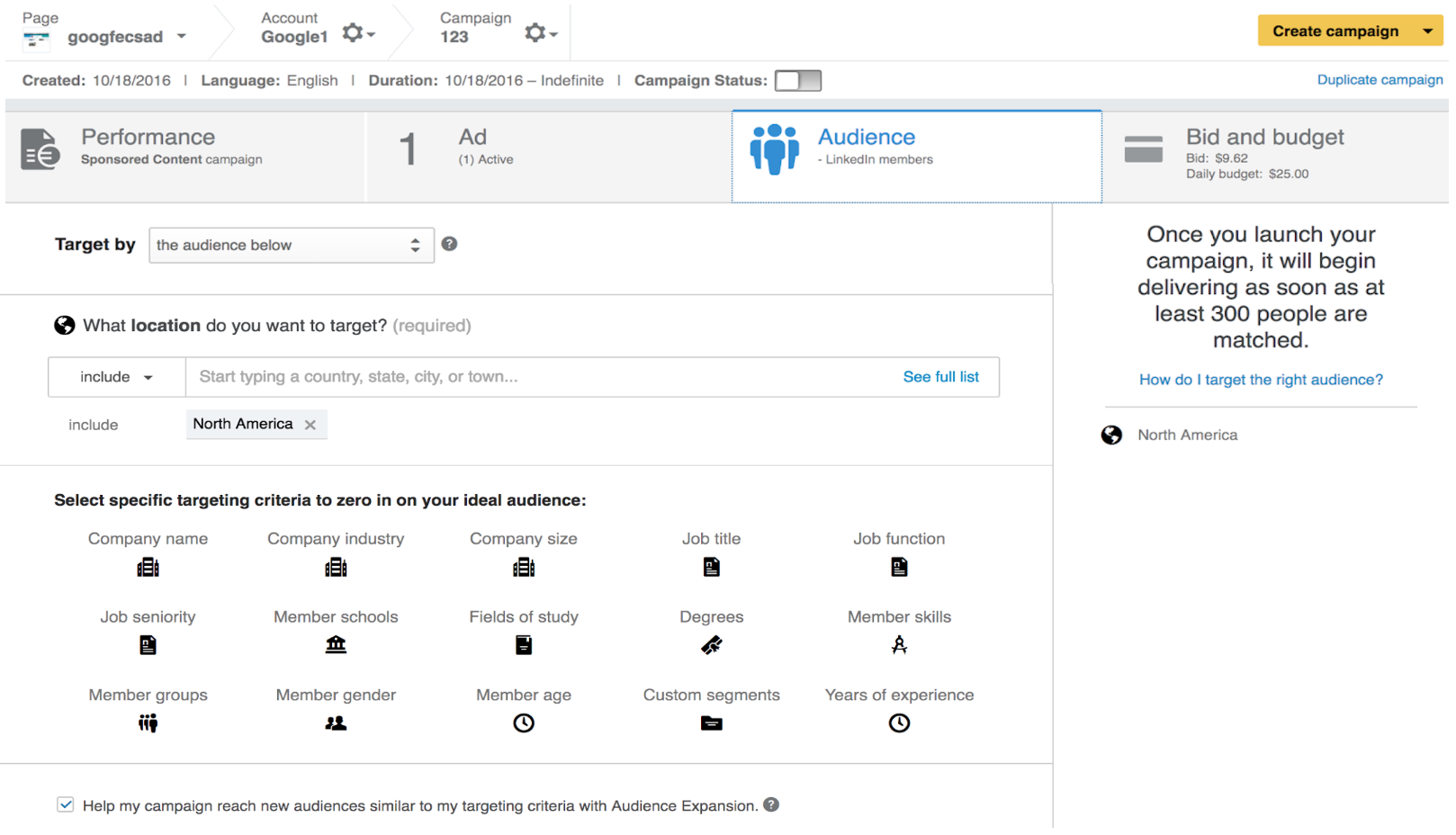}
\caption{Ad campaign and audience targeting}
\label{fig:adcampaign}
\end{figure}

{\bf Ad Targeting.}  LinkedIn Marketing Solutions (LMS) is a platform for advertisers to create ad campaigns targeting their audiences. Figure~\ref{fig:adcampaign} shows a screenshot of the ad campaign targeting interface. An advertiser can create an account on LMS, setup a campaign and then select the audience they want to reach. There can be multiple ads (`creatives') under the same campaign. To allow for flexible and effective ad targeting, LMS provides several criteria to match the audience, such as location, company size, industry, job title, and school.

\begin{figure}[t]
\centering
\includegraphics[width=0.5\textwidth]{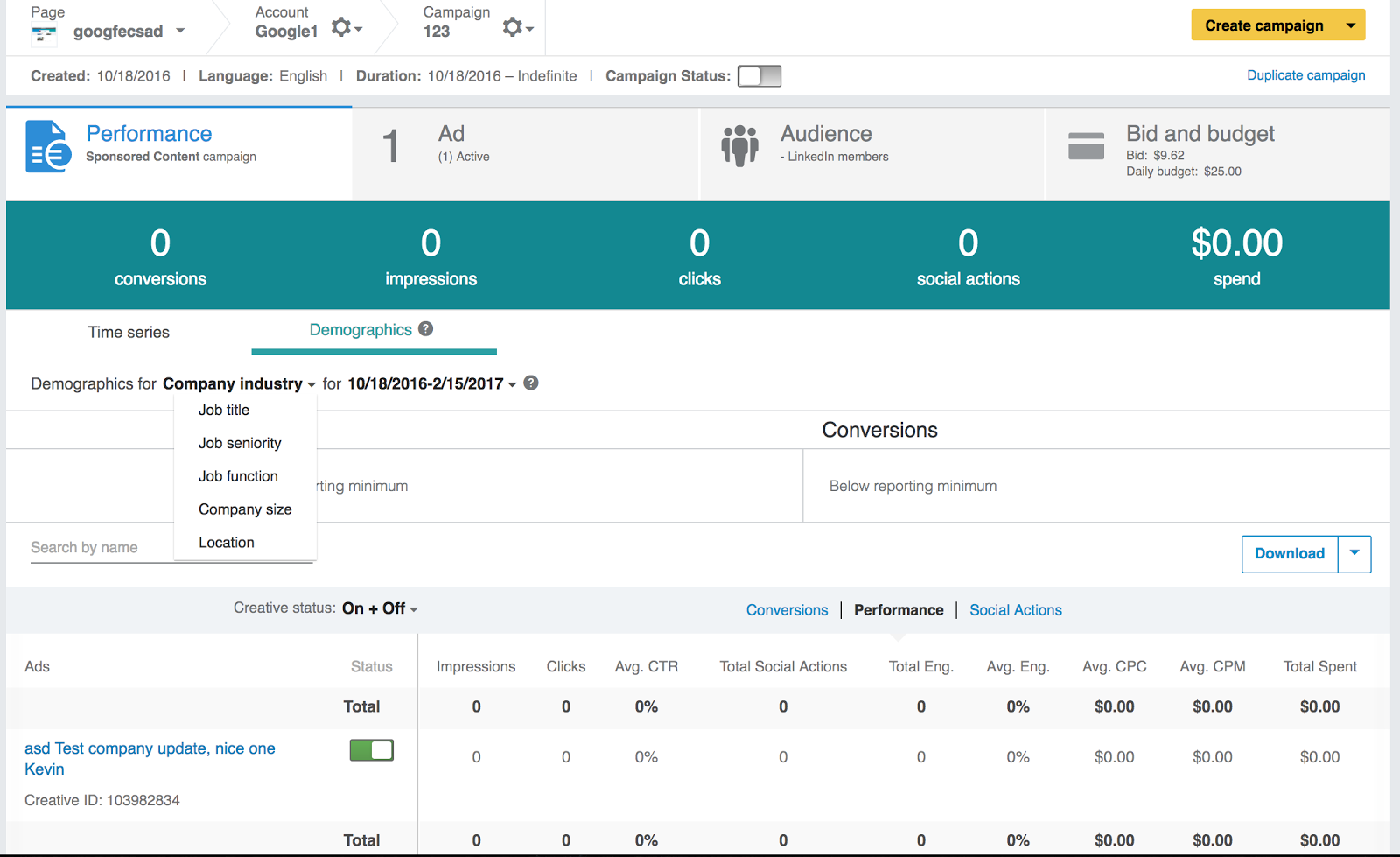}
\caption{Ad analytics}
\label{fig:adanalytics}
\end{figure}

{\bf Ad Analytics.} After setting up a campaign with one or more ads, the advertiser, or account owner, can monitor the performance of the campaign over different time periods by using the LMS Ad Analytics \& Reporting platform. The common key metrics reported include impressions, clicks, conversions, and amount spent, as shown in Figure~\ref{fig:adanalytics}. Besides viewing these metrics in aggregate for the whole campaign, the advertiser can also break them down by certain dimensions such as Job Title, Job Seniority, Function, Industry, Company, Company Size, and Location. This offers insights on the most engaged audience segments, and hence is valuable for the advertiser to understand their audience and optimize their targeting.

We deployed our privacy-preserving analytics computation system as part of LMS to prevent potential privacy attacks (\S\ref{sec:privacyrequirements}) while continuing to provide valuable insights to advertisers.

\subsection{Experimental Setup}\label{sec:expsetup}

\begin{figure*}[t]
\centering
\begin{tabular}{ccc}
\begin{subfigure}[b]{0.3\textwidth}
    \includegraphics[width=\textwidth]{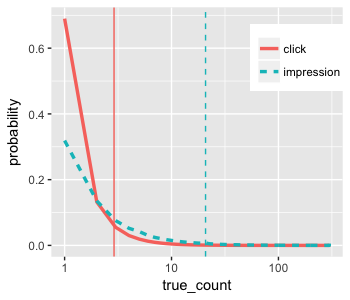}
    \caption{True impression and click distribution. Vertical lines show the means.}\label{expt:truedistributions}
\end{subfigure} &
\begin{subfigure}[b]{0.3\textwidth}
    \includegraphics[width=\textwidth]{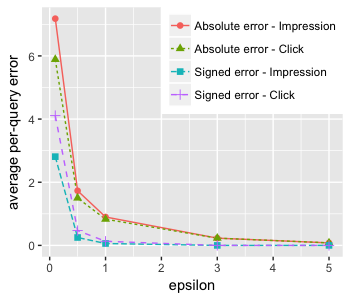}
    \caption{Absolute and signed errors with varying epsilon ($\epsilon$)}\label{expt:varyepsilon}
\end{subfigure} &
\begin{subfigure}[b]{0.3\textwidth}
    \includegraphics[width=\textwidth]{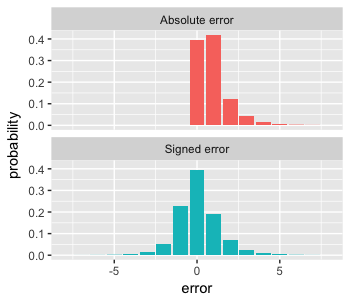}
    \caption{Distributions of added noise ($\epsilon = 1; min threshold, \minthreshold = 0$)}\label{expt:addednoise}
\end{subfigure}
\\
\begin{subfigure}[b]{0.3\textwidth}
   \includegraphics[width=\textwidth]{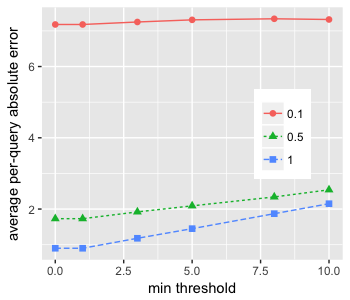}
   \caption{Absolute error with varying min threshold}\label{expt:abserror_minthreshold}
\end{subfigure} &
\begin{subfigure}[b]{0.3\textwidth}
  \includegraphics[width=\textwidth]{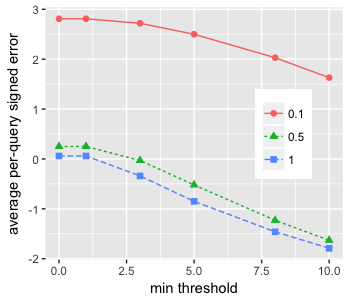}
  \caption{Signed error with varying min threshold}\label{expt:signederror_minthreshold}
\end{subfigure} &
\begin{subfigure}[b]{0.3\textwidth}
  \includegraphics[width=\textwidth]{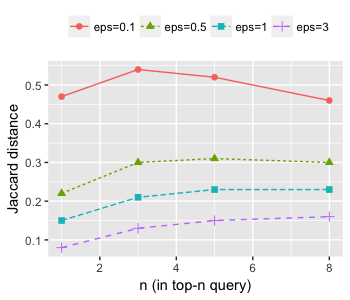}
  \caption{Jaccard distance for impression top-n queries}\label{expt:topnqueries}
\end{subfigure}
\end{tabular}
\caption{Performance evaluation with ad analytics data\label{fig:performance_eval}}
\end{figure*}

We performed our experiments using ad analytics data from December 2017.
This data contains key fields such as the granular time range, account id, campaign id, creative (ad) id, demographic attribute and value, impression counts, click counts, and conversion counts. As discussed in \S\ref{sec:arch}, this data is aggregated based on the analytics request and presented to the advertisers. The reporting platform offers both the total counts and various demographic breakdowns of the metrics. Since the total counts for metrics such as clicks and conversions can be accurately tracked by the advertisers on their site, we chose to report the true total counts, but add noise only to the demographic breakdowns. Specifically, we focus on queries for computing the number of impressions and the number of clicks for each (account, campaign, demographic category, demographic value) combination present in the dataset.\\
We first examine some statistics of the input data.
\begin{itemize}[noitemsep,leftmargin=0.3cm]
\item The daily input data has tens of millions of records on average, corresponding to several tens of thousands of accounts and several hundreds of thousands of ads (creatives).
\item Some demographic breakdowns are more granular than the others. For example, there are hundreds of thousands of possible values for Company, over 25K values for the (standardized) Job Title, but only about 10 possible values for Company Size and Job Seniority. As a result, the true counts for queries tend to vary widely across categories, e.g., impression and click counts for a specific Title are generally much lower than those for a Seniority level.
\item Figure~\ref{expt:truedistributions} shows the true distributions of daily impression and click counts for all queries returning non-zero counts. As can be seen, the distributions are long-tailed, with small medians of 1 or 2. This is because many of the demographic attribute values are very fine grained, with relatively less engagement. Note that in our application, we report only the top-10 values for each demographic attribute, so that most of the attribute values with small true counts are not surfaced to the end users. However, since these distributions resemble many other real-world distributions, we use them in their entirety to evaluate the performance of our system.
\end{itemize}

\subsection{Experiments}
We studied the performance of our system on ad analytics queries with different choices of the privacy parameter, $\epsilon$ (Expt. 1) and the minimum threshold parameter for reporting, $\minthreshold$ (Expt. 2). Then, we investigated the effect of the privacy mechanism on the top-n queries. For the first two experiments, we measured both the absolute error, $Err_{abs} = | noisy\_count - true\_count |$, and the signed error, $ Err_{sgn} = (noisy\_count - true\_count)$. The average absolute error is a measure of the overall accuracy and can be used to evaluate the utility, while the average signed error indicates how biased the reported values are (signed error = 0 means no bias). Note that the minimum threshold, $\minthreshold$ is set to 0 by default.

{\bf Expt. 1: Tradeoff between privacy and utility.}
In this experiment, we varied the privacy parameter $\epsilon$ from $0.1$ to $5$, for impression and click queries. Figure~\ref{expt:varyepsilon} shows the average absolute and signed errors vs. $\epsilon$. As expected, we observed a tradeoff between privacy and utility, with higher $\epsilon$ leading to less accurate results for both types of queries. With $\epsilon \ge 1$, the average absolute errors are less than 1.
While the average signed error is indeed less than the average absolute error as expected, we also observe that the average signed error is positive, due to the effect of setting negative noisy counts to 0 (so that we do not return negative counts).
The average signed errors reduce to almost 0 when $\epsilon\ge 1$ because the added noise then has lower variance and hence is less likely to lead to a negative noisy count.

We verified this behavior by plotting the distribution of the absolute error and the signed error for impressions with $\epsilon=1$ and $\minthreshold=0$, as shown in Figure~\ref{expt:addednoise}. We can see that the variance is small for both types of errors -- $\sim$95\% of queries have errors with magnitude at most 2.
It is interesting to see the effect of the minimum threshold on both errors. Note that if the noise added is $\ge -1$ then the noisy count is non-negative (i.e., no need to set to 0). For the absolute error, the mode is 1, instead of 0, because for queries with true count of 1 (the most prevalent) and noise added $<-1$,  the noisy count is set to 0, resulting in an absolute error of 1. For the signed error,  setting to 0 happens when $noise=-2$ and $true\ count=1$, which effectively changes the signed error from $-2$ to $-1$. Similarly, when $noise=-3$ and $true\ count=1$, the reported value is 0, effectively changing the signed error from $-3$ to $-1$. Overall there is a ``shift'' of signed error of -2 or less, towards the right-hand side, causing the noisy counts to have a small but positive bias -- the average per-query signed error in this case is 0.07.

{\bf Expt. 2: Varying minimum threshold.}
We studied the effect of varying minimum threshold $\minthreshold$ in the range from 0 to 10. Note that the value of $\minthreshold$ should be decided based on the business need in terms of the desired tradeoff between coverage and validity. While it makes sense to only report non-negative counts, in some cases, it can be desirable to report values over a certain positive threshold. Noisy counts with small magnitude have a low signal to noise ratio, and hence can cause the user to draw conclusions without sufficient validity. Note that $\minthreshold=0$ is effectively the same as $\minthreshold=1$, since the minimum positive reporting value is 1 and the rest is suppressed to 0.

We considered three values of $\epsilon \in \{0.1, 0.5, 1\}$ and computed the average absolute and signed errors for impression queries for different choices of $\minthreshold$ (shown in Figures~\ref{expt:abserror_minthreshold} and \ref{expt:signederror_minthreshold} respectively). As $\minthreshold$ is increased, the absolute errors are less sensitive for small $\epsilon$ (0.1), but vary significantly for large $\epsilon$ (1). A possible explanation is that when $\epsilon=0.1$, the magnitude of the added noise is comparable to or larger than $\minthreshold$ in the experiment range so that the effect of thresholding is masked by the noise. On the other hand, for large $\epsilon$, the noise is small, and a large minimum threshold  $\minthreshold$ could result in many small true counts (i.e., less than $\minthreshold$) to be suppressed to 0 since their noisy counts are likely to be less than $\minthreshold$.
The signed errors have the trend of decreasing as $\minthreshold$ increases (Figure~\ref{expt:signederror_minthreshold}).  This is due to the effect of suppression to 0 for most of the true counts (which mostly belong to the range [1,10]). These signed errors indicate the extent to which the reported counts are biased (0 means no bias). If unbiased results are desired, these results suggest setting $\minthreshold$ to 0 or 1 for $\epsilon=1$, to about 3 for $\epsilon=0.5$, and to even larger than 10 for $\epsilon=0.1$.

{\bf Expt. 3: Evaluating top-n queries.}
A common use case in our application is to compute top-$n$ queries for each demographic breakdown, to determine the audience segment most engaged with a given ad campaign. We hence evaluated the accuracy of results from top-$n$ queries for impressions with varying $n$. We computed Jaccard distance, defined as $1 - |A \cap B|/|A \cup B|$, where $A$ is the set of true top-$n$ results and $B$ is the set of noisy top-$n$ results, for each query, and averaged across all queries.
Note that for any query with fewer than $n$ results, we return the same whole set with or without adding noise so that Jaccard distance equals 0. About 50\% of queries in our dataset fall into this category.
Hence, we focus on queries with more than 10 values, to examine the accuracy of top-$n$ results from the noisy counts. As a result, the reported errors are larger than what we would observe in practice.

Figure~\ref{expt:topnqueries} shows the average Jaccard distance vs. $n$ for $\epsilon = 0.1, 0.5, 1, 3$. We observe that the average Jaccard distance roughly varies between 0.1 and 0.5 depending on $\epsilon$ and $n$.  As expected, Jaccard distance becomes larger as $\epsilon$ gets smaller or equivalently as noise with larger variance is added. We also observe that Jaccard distance increases when $n$ increases for larger values of $\epsilon$, but decreases with $n$ for $\epsilon = 0.1$.

\section{Lessons Learned in Practice}\label{sec:lessons}
We next present the challenges encountered and the lessons learned through the production deployment of our privacy-preserving analytics computation system as part of the LinkedIn Ad Analytics and Reporting platform for more than one year.

{\bf Business requirements and usability considerations.} 
We discuss how we took into account key business and usability factors when deploying the privacy-preserving mechanisms to the ad analytics application.

{\em Semantic consistency vs. unbiased, unrounded noise}:
Laplace mechanism for satisfying differential privacy involves noise drawn from a continuous distribution with mean of zero, so that the expectation of the noisy count would be the same as the true count. However, without any post processing, the noisy counts could be negative, while the users would expect to see cardinal numbers for counts. Although it would be ideal to add unbiased (i.e., with mean of zero) and unrounded noise to the true counts, we chose to round the noise, and also cap the reported noisy count to be at least zero, to avoid confusion for the end users.

{\em Consistency vs. utility trade-off}:
In the initial version, we computed daily noisy counts as part of an offline workflow and persisted them to storage. We then aggregated these counts whenever the analytics is requested over a multi-day range.
This implementation has advantages including simplicity of implementation, ease of debugging and maintenance, and consistency. For instance, we could retrieve the stored noisy count values for debugging any frontend issues or investigating problems reported by users. The computation cost is one-time as part of the offline workflow, instead of incurring potential latency at query time. Also, the query results would be consistent across different time ranges.
However, one main disadvantage of this approach is that the variance of the noise added increases with the time range, causing the noisy count to be less reliable for queries over a longer time range such as one month or one quarter. In our application, the users are typically interested in the performance of their ad campaigns over its lifetime rather than just on individual days. Given such usage behavior, we decided to move to an online implementation, with a hierarchical approach for time range queries (e.g., 3-hour epoch $\leftarrow$ day 
$\leftarrow$ month $\leftarrow$ quarter $\leftarrow \ldots$), trading off consistency for better utility. Further, the analytics for the current day are also computed over completed, discrete time epochs to prevent averaging attacks. For example, assume that a day is broken into 3-hour epochs, the third epoch (6am - 9am) statistics becomes available at 9:30am, and the fourth epoch (9am - 12pm) statistics becomes available at 12:15pm. Then, between 9:30am and 12:15pm, the current day analytics will make use of only the first three epochs, and hence remain unchanged.

{\em Suppression of small counts}:
The analytics and reporting applications at LinkedIn involve reporting of aggregate statistics over pre-determined and known sets of attribute values such as the set of standardized titles, the set of companies, and the set of regions. In other words, revealing the attribute value (e.g., ``Research Scientist'') as part of the reported analytics does not violate the privacy of a member. This is in contrast to applications such as releasing search queries~\cite{KKMN09}, wherein the search query itself may be private, and hence it is not desirable to reveal queries with small counts. 

However, we chose to suppress small counts for the following reason. The relative distortion due to the added noise will be large for attribute values with small counts, and hence the reported counts may not have sufficient validity for the purposes of observing patterns and making inferences.

{\bf Online computation and performance requirements. } The implementation of the online pipeline to compute noisy analytics on the fly imposes strict latency requirements. We did a few iterations to optimize the noise computation code for performance. 
For instance, we chose to implement the function, {\em GeneratePseudorandFrac} in \S\ref{sec:prln} following the approach of applying an efficient hash function and then scaling to (0, 1) range using optimized floating point computations (e.g., using double instead of BigDecimal type).
These modifications helped reduce latency significantly, resulting in a responsive user experience when interacting with the ads analytics web interface.

{\bf Scaling across analytics applications. } 
We first developed our system for preserving privacy in the context of the LinkedIn Ad Analytics and Reporting platform. In the course of refining our approaches to suit the privacy and product requirements, we chose to design our system to be broadly applicable and scalable across various analytics applications at LinkedIn. Consequently, we abstracted the application-independent functionality into a stand-alone library so that this library could be invoked as part of other analytics applications. For example, the functions for generating pseudorandom noise given query parameters are not specific to any one application, and hence are included as part of this library.

  \section{Related Work}\label{sec:related}
{\em Privacy Techniques}: Preserving user privacy is paramount when computing and releasing answers to aggregate statistical queries issued to a database containing sensitive user data. There is rich literature in the field of privacy-preserving data mining spanning different research communities (e.g.,~\cite{samarati2001protecting, Swe02, adam1989security, agrawal2000privacy, evfimievski2003limiting, kantarcioglu2004privacy, vaidya2002privacy, kenthapadi2005simulatable, aggarwal2005two, machanavajjhala2007diversity, li2007t}), as well as on the limitations of simple anonymization techniques (e.g.,~\cite{backstrom2007wherefore, narayanan2008robust, su2017anonymizing}) and on privacy attacks and challenges in ad targeting~\cite{korolova2011privacy}. Based on the lessons learned from the privacy literature, we decided to make use of the rigorous notion of differential privacy~\cite{DKM+06,DMNS06} in our problem setting. A key challenge we faced was the fact that the analytics need to be provided on a continual basis over time. Although the application of differential privacy under continual observations has been studied theoretically~\cite{CSS11,DNPR10}, we have not come across any practical implementations or applications of these techniques.

{\em Privacy Systems in Practice}: Several systems have been developed in both academic and industrial settings to address the privacy needs, especially focusing on regulations such as GDPR~\cite{kenthapadi2018privacy}. We discuss a few systems that have been either implemented or deployed in practice, and contrast them with our system.
Aircloak's Diffix is a database anonymization system designed recently to address GDPR requirements, allow a broad class of unlimited queries, and provide answers to several statistical functions with minimal distortion~\cite{francis2017diffix}. Although this system also makes use of deterministic, pseudo-random noise as in our approach, there are a few key differences. First, the queries issued in analytics application settings at LinkedIn follow a specific type of syntax corresponding to the product user interface and the exposed APIs, unlike the general class of statistical queries Diffix attempts to answer. Consequently, we could apply the rigorous notion of (event-level) differential privacy whereas Diffix does not have a formal treatment. In fact, it has been recently shown that the noise used by Diffix leaks information about the underlying data since the noise depends on the set of records that match the query conditions~\cite{gadotti2018signal}. Finally, our system has been deployed as part of LinkedIn's analytics platform and our experiments are based on real world data, while simulations are performed in Diffix~\cite{francis2017diffix}. FLEX is a system designed to enforce differential privacy for SQL queries using elastic sensitivity, and has been adopted by Uber for internal data analytics~\cite{johnson2018towards}. The key focus in this work is on computing the sensitivity (elastic sensitivity) for a broad class of SQL queries, whereas the queries allowed by our system follow a specific form, and have sensitivity of 1.
We also explored approaches similar to recent work at Google~\cite{EPK14, fanti2016building}, Apple~\cite{appleprivacy17, bassily2017practical, Gre16}, and Microsoft~\cite{ding2017collecting} on privacy-preserving data collection at scale that focuses on applications such as learning statistics about how unwanted software is hijacking users' settings in Chrome browser, discovering the usage patterns of a large number of iOS users for improving the touch keyboard, and collecting application usage statistics in Windows devices respectively. These approaches leverage local differential privacy, building on techniques such as randomized response~\cite{War65} and require response from typically hundreds of thousands of users for the results to be useful. In contrast, even the larger of our reported groups contain only a few thousand data points, and hence these approaches are not applicable in our setting.

 \section{Conclusions and Future Work}\label{sec:conclusion}
We studied the problem of computing robust, reliable, privacy-preserving analytics for web-scale applications. We presented the design and architecture of \systemname, which powers analytics applications at LinkedIn including LinkedIn Ad Analytics and Reporting platform. We highlighted unique challenges such as the simultaneous need for preserving member privacy, product coverage, utility, and data consistency, and how we addressed them in our system using mechanisms based on differential privacy. We presented the empirical tradeoffs between privacy and utility needs over a web-scale ad analytics dataset. We also discussed the design decisions and tradeoffs while building our system, and the lessons learned from more than one year of production deployment at LinkedIn. Our framework should be of broad interest for designing privacy-preserving analytics and reporting in other application settings.

A broad direction for future work would be to create a taxonomy of web-scale analytics and reporting applications, and study the applicability of different privacy approaches (e.g., interactive querying mechanisms vs. data perturbation and publishing methods) for each class of applications in the taxonomy. Another direction would be to associate a notion of utility with the availability / correctness of different types of statistics, and to formulate as a utility maximization problem given constraints on the `privacy loss budget' per user. For example, we could explore adding more noise (i.e., noise with larger variance) to impressions but less noise to clicks (or conversions) since the number of impressions is at least an order of magnitude larger than say, the number of clicks. Similarly, we could explore adding more noise to broader time range sub-queries and less noise to granular time range sub-queries towards maximizing utility.

\begin{acks}
The authors would like to thank all members of LinkedIn Ad Analytics and Reporting team for their collaboration for deploying our system as part of the launched product, and
Deepak Agarwal,
Mark Dietz,
Taylor Greason,
Sara Harrington,
Ian Koeppe,
Sharon Lee,
Igor Perisic,
Rohit Pitke,
and
Arun Swami
for insightful feedback and discussions.
\end{acks}

{
\bibliographystyle{abbrv}  \bibliography{paper}
}

\end{document}